\documentclass[a4paper]{article}
\usepackage[a4paper, left=1.2in, right=1.2in, top=1in, bottom=1in]{geometry}
\usepackage{graphicx}
\usepackage{amsmath,amssymb}
\usepackage{mathtools,xcolor}
\usepackage{float}
\usepackage{authblk}
\usepackage{bm}

\usepackage{cite}
\usepackage[nottoc]{tocbibind}
\def\rl{\longleftrightarrow}

\newcommand{\la}{\langle}
\newcommand{\ra}{\rangle}

\newcommand{\be}{\begin{equation}}
	\newcommand{\ee}{\end{equation}}
\newcommand{\ba}{\begin{eqnarray}}
	\newcommand{\ea}{\end{eqnarray}}

\def\h{\hskip 1cm}

\def\lo{\longrightarrow}
\def\a{\alpha}
\def\b{\beta}

\def\ni{\noindent}

\title{\bf A review of perfect quantum state transfer, from one to two and three dimensional arrays  of qubits. }

\author[1]{Marzieh Asoudeh \footnote{Corresponding Author, email: marzieh.asoudeh@gmail.com}}
\affil[1]{Department of Physics, NT.C., Islamic Azad University, Tehran, Iran.}

\author[2]{Vahid  Karimipour \footnote{email: vahid.karimipour@gmail.com}}
\affil[2]{Department of Physics, Sharif University of Technology, Tehran, Iran}

\date{}

\begin{document}
	\maketitle

	\begin{abstract}
		In the light of recent advances in fabricating single layer quantum chips and a possible road toward development of multi-layer quantum chips, we review, in a detailed way, the subject of quantum state transfer with particular emphasis on perfect quantum state transfer in two and three dimensional lattices. We show how one can route an unknown quantum state from one node in a single layer of a quantum chip to another one on another layer with unit fidelity. 
		 Our method of presentation in this review  allows the reader with a modest background in quantum mechanics to grasp the essential ideas and methods of this important branch of quantum information theory. 
		\bigskip
	\end{abstract}

	\maketitle

	\section{Introduction}
	Recent years have witnessed significant progress in the development of quantum integrated circuits and quantum chips  and a breakthrough in proving supremacy of quantum computers over classical ones \cite{sycamore}.  Advances in qubit coherence times, gate fidelities, and quantum error correction have enabled the scaling of quantum processors, pushing the boundaries of what can be achieved in quantum computation and communication. Key technologies, specfically construction of arrays of Josephson junctions and superconducting qubits and miniaturization of these arrays into small chips, has made it possible  the development of increasingly complex quantum architectures.  As these technologies mature, efficiently transferring quantum information across chips becomes a central challenge. \\
	
	Currently, multi-layered structures are widely used in classical integrated chips. In fact current integrated chips contains even up to 10 layers which consist of  device layers, dielectric layers and metal layers to prevent short circuits. 	
	If such multilayer chip architectures are adapted for quantum technology, it becomes crucial to develop reliable techniques for transferring quantum states in 3D structures. to develop techniques for efficient and reliable transfer of quantum states in three dimensional layered structures. \\
	
	The fundamental problem of  quantum state transfer, the ability to use the natural dynamics of quantum spin chains to reliably transfer a quantum state from one end to the other has a long and fascinating history. Quite interestingly, the interaction between Josephson junctions gives rise to terms that resemble spin-spin interactions, typically of the XY type, which in the context of quantum state transfer has been the main model of study. The idea of state transfer between two ends of a one-dimensional array of interacting spins interacting was originated in \cite{bosePRL} and then developed in numerous other works \cite{ekert, Yung, vinet1, vinet2,  Karbach, Franco, Gualdi, Markiewicz}, \cite{Shi, Plenio, Wojcik, Burgarth1, Franco2, Feldman,Yao} and \cite{Burgarth1, Osborne, Burgarth2,Cappellaro,Zhang, Wichterich, Wang, Banchi}. Many new ideas and modifications were taken into account, including but not restricted to  the effect of imperfections on the couplings \cite{Zwick}, thermal fluctuations \cite{bayatmethermal},  transfer of  higher dimensional states or qudits (\cite{bayatme}, \cite{asoudehme}) and experimental realizations in different platforms \cite{chapman, chapman2, Dolev, exp1, exp2, exp3, Axline}. Of particular interest to us in this review, is a trend on Perfect State Transfer (PST) which originated in the work of Christandl et al \cite{ekert}. The idea of this paper was to engineer the couplings between the various spins in the chain in such a way that
	states are transferred with perfect  fidelity. This idea was then developed in various works,  \cite{ekert, Yung, vinet1,vinet2, general1, Karbach, Franco, Gualdi, Markiewicz} or with
	arbitrary high fidelity \cite{Shi, Plenio, Wojcik, Burgarth1, Franco2,Feldman,Yao, lorenzo1, lorenzo2, bayatd}. In addition, some minimal external control on the chain
	dynamics was also introduced in order to
	achieve similar results \cite{ Burgarth3, Osborne, vinet3 , Burgarth2,Cappellaro,Zhang, Wichterich, Wang, Banchi, Kay}. \\
	
	In particular in \cite{Burgarth2}, the authors 
	presents a method, called conclusive state transfer,  for achieving arbitrarily perfect quantum state transfer through dual-rail encoding, i.e. where  a logical qubit state $\alpha|0\ra+\beta |1\ra$ is encoded into the state to two physical qubit $\alpha |1,0\ra+\beta |0,1\ra$ which is transferred through two quantum chains. It is shown in \cite{Burgarth2} that  even when the two quantum chains used for communication have random coupling fluctuations and are not identical. The authors show that perfect transfer is possible not only in the presence of mild randomness (up to 5 percent noise) but also without requiring precise knowledge or control over the chain's internal structure. They propose a method in which only the ends of the chain, accessible to the sender (Alice) and the receiver (Bob), need to be manipulated, while the rest of the chain acts like a black box. Importantly, the receiver (Bob) doesn't need to know the full Hamiltonian—he only needs to measure transition amplitudes at the chain ends and search numerically for optimal times. However, one should note that while the fidelity in this protocol can be made arbitrarily close to unity, the probability of successful transfer at any given measurement time is less than one. Therefore, the receiver should make multiple measurements at different times, to successfully retrieve the state. In fact the dynamics of the chain spreads the state over many modes in the disordered chains.
	The probability amplitude for the correct output state (needed for perfect transfer) oscillates over time and is not always sharply peaked.
	Bob performs measurements when the transfer amplitude peaks. If a measurement fails, the system evolves unitarily until the amplitude returns, offering another opportunity for success. We will briefly discuss this method in section (\ref{conclusive}). \\
		
	This basic idea was then generalized to multiple chains 
	in \cite{Burgarth2}, where 	
	perfect fidelity and optimal efficiency is achieved by using multiple parallel, non-interacting quantum chains. The sender prepares collective excitation states across a subset of 
	$M$
	chains, and the receiver performs joint measurements at the opposite ends. The authors prove that, under a mild and easily satisfiable condition on the system's Hamiltonian, the probability of successful state transfer converges to unity as the measurement procedure is repeated. Furthermore, the encoding achieves asymptotic communication efficiency 
	$R
	=
	1$
	, meaning that one logical qubit can be transmitted per physical chain.\\
	
	In this review, out of the multitude of works on PST, we focus on one particular line of research which starts from the seminal work of \cite{ekert} on engineered non-uniform couplings and passes through the refinement proposed in \cite{Kay, KayNN} on quasi-one dimensional chains with almost uniform couplings to the work of the authors on generalization of PST to two and three dimensional lattices. Thus, we have to leave out  many other works which are outside of this particular line of research, e.g. PST using quantum random walks \cite{Zhan} or PST on networks with PT-symmetric  non-Hermitian networks \cite{expZhang}. For a review see \cite{bayatbook}.\\

	The structure of this paper is as follows. After reminding the basic idea of state transfer and setting up our notation in section (\ref{basicrem}),  we explain the basic idea of quantum state transfer \cite{bosePRL} in (\ref{Qst}) according to which such a transfer is possible with high, but not perfect fidelity.  A natural question is then whether or not quantum state transfer is possible with perfect fidelity.  We will then briefly discuss the work of \cite{Burgarth1} on conclusive state transfer, which requires multiple measurements at the destination and a probability of success which tends to unity after many measurements. The question of perfect state transfer with specific timing, is taken up in  section (\ref{pstperfect}), where we review the seminal work of  \cite{ekert} in which they show with engineered non-uniform couplings perfect state transfer is possible.  In section (\ref{pstkay}), we show how the undesirable feature of non-uniform couplings can be remedied at the price of adding global control to the chain \cite{Kay}.   Finally in section (\ref{pstme}) we present our own method for perfect state transfer in two dimensional and three dimensional arrays of spins. In this section we add many details which are absent in the original paper \cite{sarmadi} and present our scheme in a detailed way in order to make it accessible to a wide readership. The paper ends with a discussion.\\

	\section{Preliminaries} \label{basicrem}
		The Heisenberg spin system is a cornerstone of condensed matter physics, playing a crucial role in our understanding of magnetism and strongly correlated systems. Originating from Werner Heisenberg's pioneering work in the 1920s, the Heisenberg model describes interactions between neighboring spins in a magnetic material. This model is particularly significant because it captures the essence of quantum mechanical interactions in magnetic systems, where the spin of electrons leads to a variety of magnetic phenomena.	
	In the Heisenberg model, the Hamiltonian typically takes the form 
		\be
	H=-J\sum_{\la i, j\ra} X_iX_j+Y_iY_j+Z_iZ_j-B\sum_i Z_i,
	\ee
	where $X_i, Y_i$ and $Z_i$ are the Pauli operators respectivley acting as 
	$X=\begin{pmatrix}0&1\\ 1&0\end{pmatrix},\ \ Y=\begin{pmatrix}0&-i\\ i&0\end{pmatrix},\ $ and $X=\begin{pmatrix}1&0\\ 0&-1\end{pmatrix}$ on site $i$, 
		 $B$ is the magnetic field (which we take to be positive),
	$J $
	 is the exchange interaction parameter, and $\la i, j\ra$ 
 denotes summation over nearest-neighbor pairs.  Depending on the sign of 
	$J$ the system can exhibit ferromagnetic (
	$J>0$) or antiferromagnetic ($
	J
	<
	0$
	) order. This model provides a simplified yet powerful framework to study the collective behavior of spins and the resulting magnetic properties of materials.\\

	The Hamiltonian of the Heisenberg spin chain  has the
	following symmetries
	\begin{equation}
		[H,Z_{tot}]=0,
	\end{equation}
	where
	\begin{equation}\label{Js}
		Z_{tot}:= \sum_{i=0}^{N-1} \sigma_{z,i},
	\end{equation}
	is the third component of the total spin.
	Here $|s_i\ra, \ \  s_i = \pm 1$ is the state of the $i$-th
	spin expressed in terms of the eigenvectors of $\sigma^z$, i.e.
	$\sigma^z_i|s_i\ra=s_i|s_i\ra$. 	In accordance with the notation used in quantum computation, we denote the two eigenstates of the $\sigma_z$ operator, namely $|+\ra$ and $|-\ra$ respectively by $|0\ra$ and $|1\ra$. Thus the operator $S_z$ counts the number of up spins minus the number of down spins.
	In a strong magnetic field, the ground state of the system is 
	\be\label{gs}
	|\psi_0\ra=|0,0,0\cdots 0\ra.
	\ee
	The full spectrum of the model in one dimension can be found by using the Bethe Ansatz. However the low-lying states, i.e. the states in the one-particle sector can easily be found. 	
 The first excited states are linear combination of one
	particle states
	\begin{equation}\label{1par}
		|\psi\ra = \sum_{k}c_k |k\ra,
	\end{equation}
	where
	\begin{equation}\label{k1}
		|k\ra := |0,0,0,\cdots ,1,\cdots ,0,0,0\ra,
	\end{equation}
	in which only the spin in the $k-$th site is down. \\
	\subsection{Notations and Conventions}\label{not}
The computational basis states are given by 
	\be
	|0\ra=\begin{pmatrix}1\\ 0\end{pmatrix},\ \ \ \ |1\ra=\begin{pmatrix}0\\ 1\end{pmatrix}.
	\ee
	thus we have
	\be
	X|0\ra=|1\ra,\ \ \ X|1\ra=|0\ra,\ \ \ \ \ \ \ Y|0\ra=i|1\ra,\ \ \ Y|1\ra=-i|0\ra \ \ \ \ Z|0\ra=|0\ra,\ \ \ Z|1\ra=-Z|1\ra.
	\ee

\ni	A state in which only the $k-$th spin is excited is denoted by $|{ k}\ra$, or 
	\be
	| k\ra=|0,0,0\cdots 1,\cdots 0,0,0\ra.
	\ee
	Since there are different types of qubits in the lattices which we will discuss below, we use two different words for two dimensional lattices. A layer of qubits accommodates the data qubits and planes of qubits (3 of them) are the control planes which are acted by regularly timed interval of global pulses for routing the data qubits. 	
	
		\section{The basic idea of quantum state transfer}\label{Qst}
	 Quantum registers are usually envisaged as an array of qubits on which different single-qubit ot double-qubit gates (like CNOT gate) can act and evolve this state. No matter how physical qubits are constructed, being a sequence of charged ions in an ion trap, or a sequence josephson junctions, they can always be modelled by two-level spin states which interact via a specificly tuned exchange interaction and are also under the influence of an external magnetic field. A basic task in any quantum computer or quantum simulator is to transfer a quantum state from one register to another. This may be within a single quantum computer or between different quantum computers. While photons are the ideal carriers
	 of quantum information over long distances, it has become evident
	 that the best possible method for transferring quantum
	 information over short distances, i.e. through regular arrays of
	 qubits, is to exploit the natural dynamics of the many body
	 system. This idea was first introduced in the work of Bose
	 \cite{bosePRL} who showed that the natural dynamics of a
	 Heisenberg ferromagnetic chain can achieve high-fidelity transfer
	 of qubits over distances as long as 80 lattices units. 	
	 The basic idea can explained as follows. Consider the chain with the pattern of couplings shown in figure (\ref{nonuniform})

	 \begin{figure}[H]
	 	\centering
	 	\includegraphics[width=1\columnwidth]{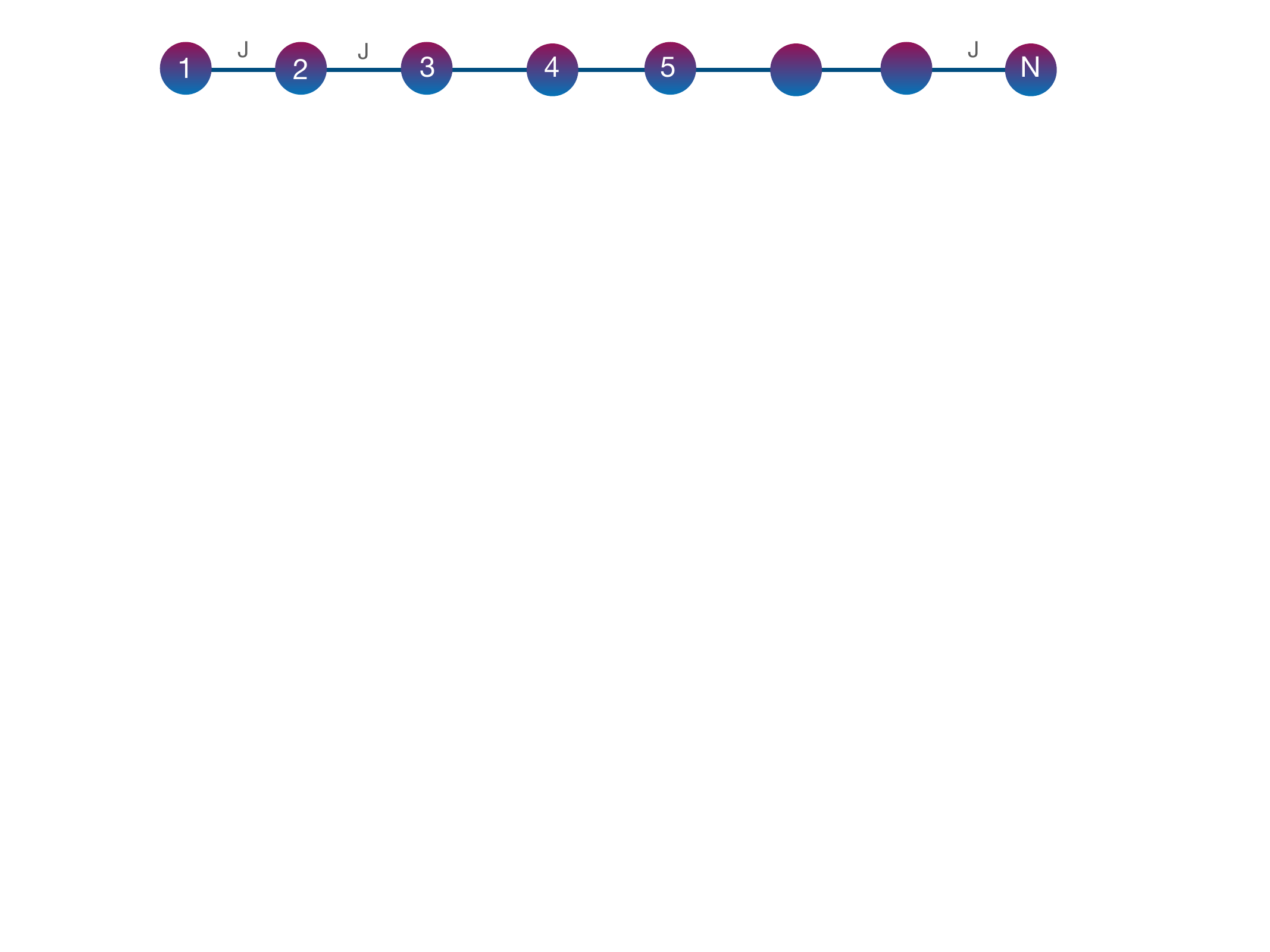}\vspace{-8cm}
	 	\caption{The quasi-one dimensional chain with uniform couplings for perfect transfer of quantum states. }
	 	\label{basic}
	 \end{figure}

	 The Hamiltonian governing the interaction of the spin in the
	 channel is given by 
	 \begin{equation}\label{Heis2}
	 	H = -J\sum_{i=1}^{N-1} {\bf \sigma}_i\cdot{\bf \sigma}_{i+1} +
	 	B \sum_{i=0}^{N} \sigma_{z,i},
	 \end{equation}
 The ground state of this chain is nothing but 
 \be
 |gs\ra=|0,0,0,\cdots 0\ra
 \ee
 where $|0\ra$ stands for a spin up state in the $z$ direction. This state is therefore invariant under the dynamics of the Hamiltonian In order to transfer an arbitrary state $|\phi\ra_1=(a|0\ra+b|1\ra)_1$ which at time $0$ resides at site $1$ to the end of the chain, 
 we need to coherently transfer the state $|1\ra_1$ to the end of the chain. This means that the Hamiltonian is expected to do the following:
 \be
 e^{-iHT}|1,0,0,\cdots\ra=|0,0,0,\cdots 1\ra,
 \ee
 if we stop the dynamics at a suitable time $T$. The problem is that this does not happen in general and the Hamiltonian evolution does not take initial state to the right hand side at any time, but takes it to a superposition of one-particle states in the form
 \be
 e^{-iHT}|1,0,0,\cdots 0\ra=\sum_{j=1}^N  |{\bf e}_j\ra\la {\bf e}_j|e^{-iHT}|1,0,0,,\cdots \ra,
 \ee
 where $|E_\a$'s are the one-particle sector of the energy eigenstates of $H$. The higher particle sectors do not contribute to the sum on the right hand side. This means that a general state $|\phi\ra_1|0,0,\cdots\ra_{2,\cdots N}$, where $$|\phi\ra=\cos\frac{\theta}{2}|0\ra+e^{i\phi}\sin\frac{\theta}{2}|1\ra$$  transforms to 
 \be
 |\Psi(t)\ra=\cos\frac{\theta}{2}|0,0,0,\cdots 0\ra+e^{i\phi}\sin\frac{\theta}{2}\sum_{j=1}^N|{\bf e}_j\ra\la {\bf e}_j|e^{-iHT}|1,0,0,\cdots 0\ra .
 \ee
 This shows that the state of the last qubit is not a pure state, but a mixed one given by 
 \be
 \rho_N=P(T)|\psi(T)\ra\la \psi(T)|+(1-P(T))|0\ra\la 0|,
 \ee
 where
 \be
 |\psi(T)\ra=\frac{1}{\sqrt{P(T)}}(\cos\frac{\theta}{2}|0\ra+e^{i\phi}\sin\frac{\theta}{2}g_N(T)|1\ra) \ee
 in which
 \be
 P(T)=\cos^2\frac{\theta}{2}+\sin^2\frac{\theta}{2}|g_N(T)|^2,\h g_N(T)=\la {\bf e}_N|e^{-iHT}|1\ra
 \ee
 One can then calculate the fidelity of this state with the initial state and average it over the unit sphere to find the average fidelity of state transfer
 \be
 \overline{F}:=\frac{1}{4\pi}\int\la \phi|\rho_N|\phi\ra d\phi.
 \ee
 A straightforward calculation shows that 
 \be
 \overline{F}=\frac{1}{3}|g_N(T)|\gamma+\frac{1}{6}|g_N(T)|^2+\frac{1}{2},
 \ee
 in which $\gamma=arg(g_N(T)).$
 It was then shown in \cite{bosePRL} that in this way the average fidelity of state transfer is higher than $\frac{2}{3}$, for chains of length up to $N=80$.\\
 	 	 
 	 	 \section{Conclusive state transfer} \label{conclusive}
\begin{figure}[H]
 	 	 	\centering
 	 	 	\includegraphics[width=1\columnwidth]{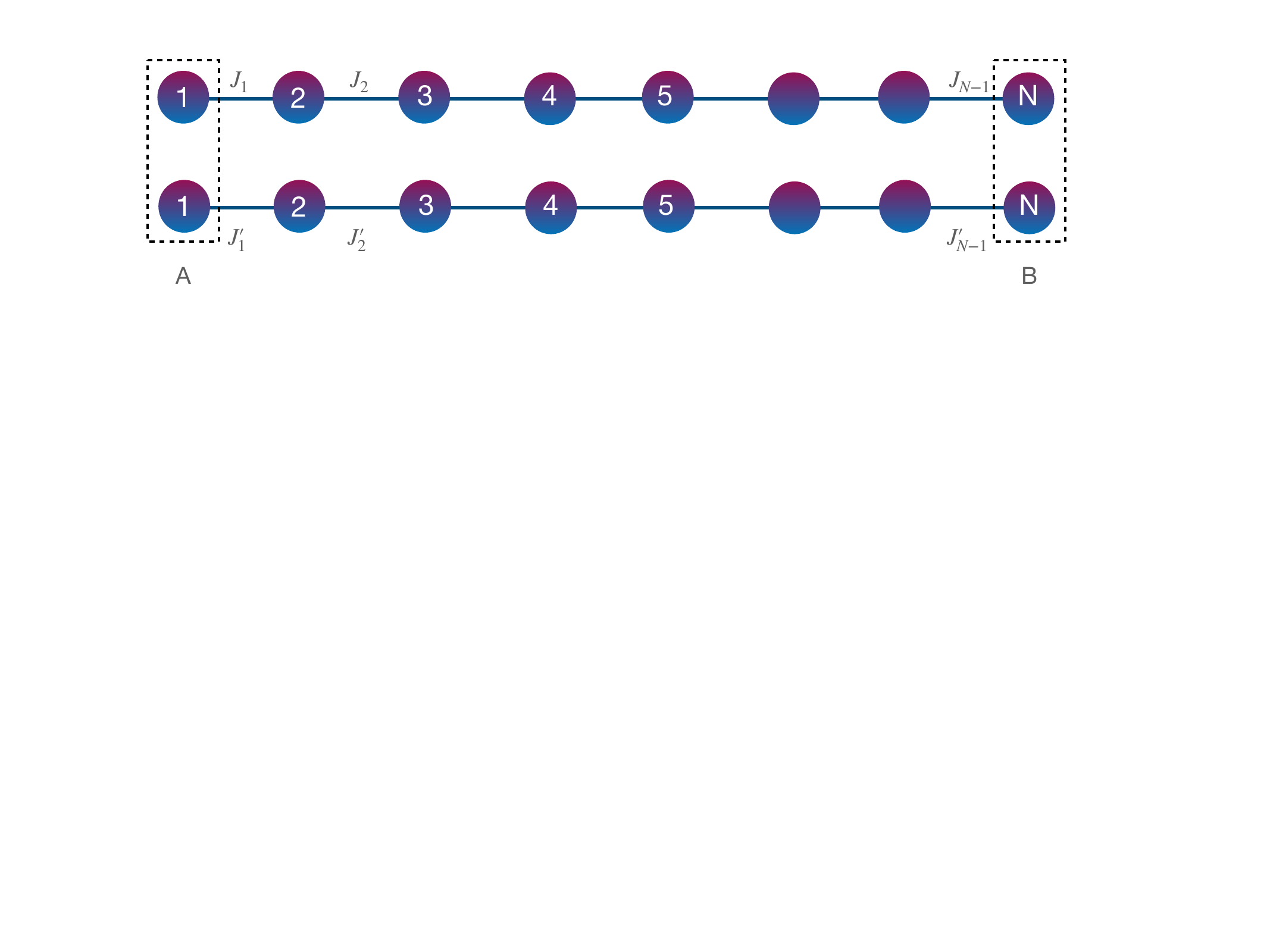}\vspace{-8cm}
 	 	 	\caption{The double chain setup for conclusive state transfer of \cite{Burgarth1,Burgarth2}. The first two qubits are controlled by Alice (A) and the last two ones by Bob (B).The protogol is robust if the couplings show small variations, i.e. of the form $J_i=J(1+\delta_i)$ and $J'_i=J(1+\delta'_i)$ for small $\delta_i$ and $\delta'_i.$.  }
 	 	 	\label{Basic2}
 	 	 \end{figure}

 	 	 In this section we briefly discuss the work of \cite{Burgarth1,Burgarth2}, who propose a method in which the receiver Bob, can finally retrieve the state after a number of measurements. They show that a  single spin-1/2 quantum chain could not be used for conclusive transfer, instead they use a dual rail encoding of state on two chains (figure (\ref{Basic2})). 

 	\ni 	 That is, Alice encodes her state 
 	 	 \be
 	 	 \a|0\ra+\beta|1\ra\lo \a|1,0\ra+\beta |0,1\ra
 	 	 \ee
 	 	 and insert it on the left end of two parallel chains. The chains are described by the two Hamiltonians $H^{(1)}$ and $H^{(2)}$ acting separately on the two chains. The total Hamiltonian is thus
 	 	 \begin{equation}
 	 	 	H = H^{(1)} \otimes I^{(2)} + I^{(1)} \otimes H^{(2)},
 	 	 \end{equation}
 	 	 and the time evolution operator is given by
 	 	 \begin{equation}
 	 	 	U(t) = e^{-i H t} = e^{-i H^{(1)} t} \otimes e^{-i H^{(2)} t}.
 	 	 \end{equation}
 	 	 
 	 	 It is assumed that both chains have equal length $N$, an assumption that we also use in the following description, but was shown not to be essential in the proposal of \cite{Burgarth1, Burgarth2}. Suppose the Hamiltonians commute with the total $z$-component spin:
 	 	 \begin{equation}
 	 	 	\left[H^{(i)}, \sum_{j=1}^N \sigma_z^{(i)}\right] = 0.
 	 	 \end{equation}
 	 	  	 	 Then the vacuum state $|0\rangle^{(i)} = |00\cdots0\rangle^{(i)}$ is an eigenstate, and states with a single excitation take the form
 	 	 \begin{equation}
 	 	 	|m\rangle^{(i)} = |00\cdots 1_m \cdots 0\rangle^{(i)}, \quad 1 \le m \le N.
 	 	 \end{equation}
 	 	 	 	 States of the total Hilbert space are denoted as
 	 	 \begin{equation}
 	 	 	|m,n\rangle = |m\rangle^{(1)} \otimes |n\rangle^{(2)}.
 	 	 \end{equation}
 	  	 	 Alice controls the first qubit and Bob controls the last qubit of each chain. The inserted state of Alice evolves as , 
 	 	 \begin{equation}
 	 	 	\sum_{n=1}^N \left\{ \alpha g_{n,1}(t) |0,n\rangle + \beta f_{n,1}(t) |n,0\rangle \right\},
 	 	 \end{equation}
 	 	 where
 	 	 \begin{align}
 	 	 	f_{n,1}(t) &= \langle n,0 | U(t) | 1,0 \rangle, \\
 	 	 	g_{n,1}(t) &= \langle 0,n | U(t) | 0,1 \rangle.
 	 	 \end{align}
 	  	 	 At some time $t_1$, where the proportionality holds,
 	 	 \begin{equation}
 	 	 	g_{N,1}(t_1) = e^{i \phi_1} f_{N,1}(t_1),
 	 	 \end{equation}
 	 	 the state becomes
 	 	 \begin{equation}
 	 	 	\sum_{n=1}^{N-1} \left\{ \alpha g_{n,1}(t_1) |0,n\rangle + \beta f_{n,1}(t_1) |n,0\rangle \right\} + f_{N,1}(t_1) \left( e^{i\phi_1} \alpha |0,N\rangle + \beta |N,0\rangle \right).
 	 	 \end{equation}
 	  	 	 Bob applies a CNOT gate  and measures the second qubit. If the outcome is $|N\rangle^{(2)}$, the state is transferred up to a known phase which they can correct. If not, the process can be repeated, until he finally retrieve the state with high enough fidelity. The advantage of this protocol is that it is robust against imperfections in the couplings of the spin. The price which is paid for this important and highly desirable advantage, is that Bob has to  perform a number of measurements to retrieve the state successfully. Depending on the amount of disorder in the chains, the number of measurements ranges between 20 and 120 \cite{Burgarth2}. 
 	 	 
 \section{Perfect State Transfer with definite timing}\label{pstperfect}	 	 	 
 	Since the proposal of S. Bose \cite{bosePRL} for transferring quantum
	states via natural evolution of quantum spin one-half chains, this idea was extended in many different directions. In particular it was asked how one can transfer states with perfect fidelity, i.e. in the same exact form in a predetermined time, in contrast with the work of \cite{Burgarth1,Burgarth2}. To this end, we first describe Perfect State Transfer (PST) on short chains of length two and three which will later play a crucial role in the proposal for PST on  quasi-one dimensional chains \cite{Kay} and two and three dimensional lattices \cite{sarmadi}.

	\subsection{Perfect State Transfer in chains of length 2 and 3}
	Let us see why the Heisenberg Hamiltonian can perfectly transfer quantum states in chains of length 2 and 3. The Hamiltonian for chain of length 2 is given by
		
	\be
	H=\frac{1}{2}(X\otimes X+Y\otimes Y)
	\ee
	from which we obtain
	\be
	H|0,0\ra=H|1,1\ra=0,\h H|1,0\ra=|0,1\ra,\ \ \ H|0,1\ra=|1,0\ra 
	\ee
	or in matrix form in the one-particle sector 
	\be
	H=\begin{pmatrix}0&1\\1&0\end{pmatrix}.
	\ee
	This means that 	
	\be
	U(T)=e^{-iHT}=\cos T I-i\sin T H
	\ee
	which implies that 
	\be
	U(T=\frac{\pi}{2})=-iH
	\ee
	Thus we find	
	\be
	U(\frac{\pi}{2})|1,0\ra=-i|0,1\ra.
	\ee
This means that in a time of $T=\frac{\pi}{2}$, the evolution operator transfers an excitation from one side of the chain to the other in a perfect manner. Therefore, given an arbitrary state $|\phi\ra=\a|0\ra+\beta |1\ra$, it is transferred with an extra phase,
\be
U:\ \ (\a|0\ra+\beta |1\ra)|0\ra\lo |0\ra(\a|0\ra-i\beta |1\ra).
\ee
Given the fact that this phase is known, it can be readily removed at the right hand side by the action of a correcting operator $Y$. Consider now a chain of length $3$.
Consider now a chain of length 3. The Hamiltonian is  
	\be
	H=P_{12}+P_{23}
	\ee
	and its action on the one-particle sector is 
	\ba
	H|1,0,0\ra&=&|0,1,0\ra\cr
	H|0,1,0\ra&=&|1,0,0\ra+|0,0,1\ra\cr
	H|0,0,1\ra&=&|0,1,0\ra.
	\ea
	In matrix form the Hamiltonian takes the form
		
	\be
	H=\begin{pmatrix}0&1&0\\1&0&1\\ 0&1&0\end{pmatrix}
	\ee
	where its square is 
	\be
	H^2=\begin{pmatrix}1&0&1\\0&2&0\\ 1&0&1\end{pmatrix}.
	\ee
	It is then readily seen that.
	\be
	H^3=2H.
	\ee
	This implies that the eigenvalues of $H$ are $\{0,\sqrt{2},-\sqrt{2}\}$ which allows us to determine the evolution operator as 
	\be
	U=e^{-iHT}=\a I+\beta H+\gamma H^2.
	\ee
	Matching of the eigenvalues of both sides leads to the final form of the evolution operator 
	\be
	U(T)=e^{-iHT}=I-\frac{i}{\sqrt{2}}\sin(\sqrt{2}T)H+\frac{\cos(\sqrt{2}T)-1}{2}H^2.
	\ee
	At time $T=\frac{\pi}{\sqrt{2}}$, the evolution operator in matrix form will be 
	\be
	U(T=\frac{\pi}{\sqrt{2}})=I-H^2=\begin{pmatrix}0&0&-1\\0&-1&0\\ -1&0&0\end{pmatrix}
	\ee
	which is seen to be in mirror form and transfers a state $|1,0,0\ra$ to $|0,0,1\ra$ modulo a phase. 
	\be
	U(T=\frac{\pi}{\sqrt{2}})|1,0,0\ra=-|0,0,1\ra.
	\ee
	Thus the evolution operator transfers an arbitrary state $|\phi\ra=\a|0\ra+\beta |1\ra$ from the left hand side of the chain to the right hand side in the following form 
	\be
	U(T=\frac{\pi}{\sqrt{2}}):(\a|0\ra+\beta|1\ra)|0,0\ra\lo|0,0\ra(\a|0\ra-\beta|1\ra).
	\ee
	which can then be corrected by an appropriate phase gage $Z$.
	
	\subsection{Perfect State Transfer in chains of arbitrary length}\label{PSTN}
\begin{figure}[H]
		\centering
		\includegraphics[width=13cm]{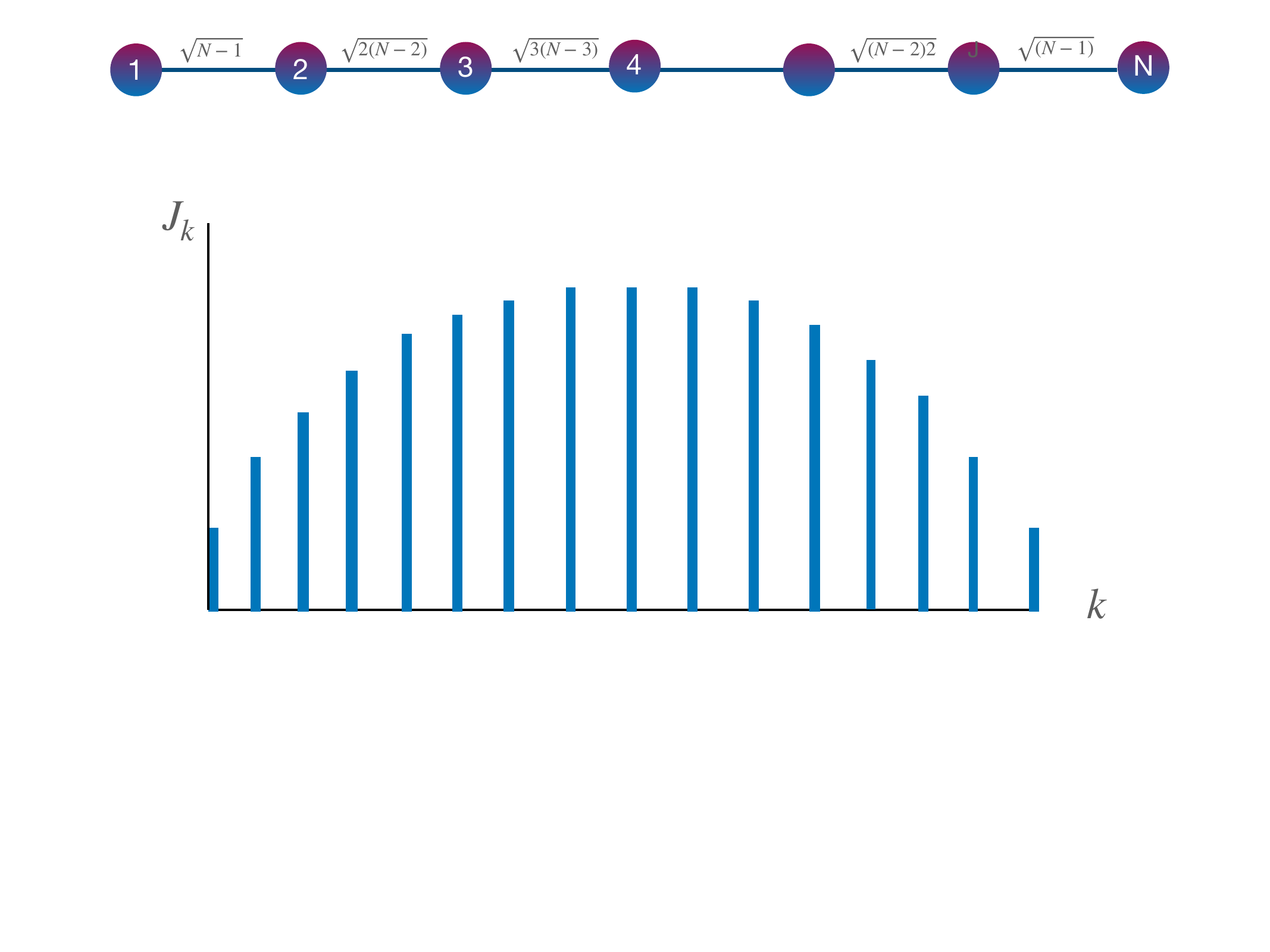}\vspace{-3cm}
		\caption{The pattern of couplings in a short chain. The barchart below shows the couplings for a chain of length 20. }
		\label{nonuniform}
	\end{figure}

\begin{figure}[H]
	\centering
	\includegraphics[width=15cm]{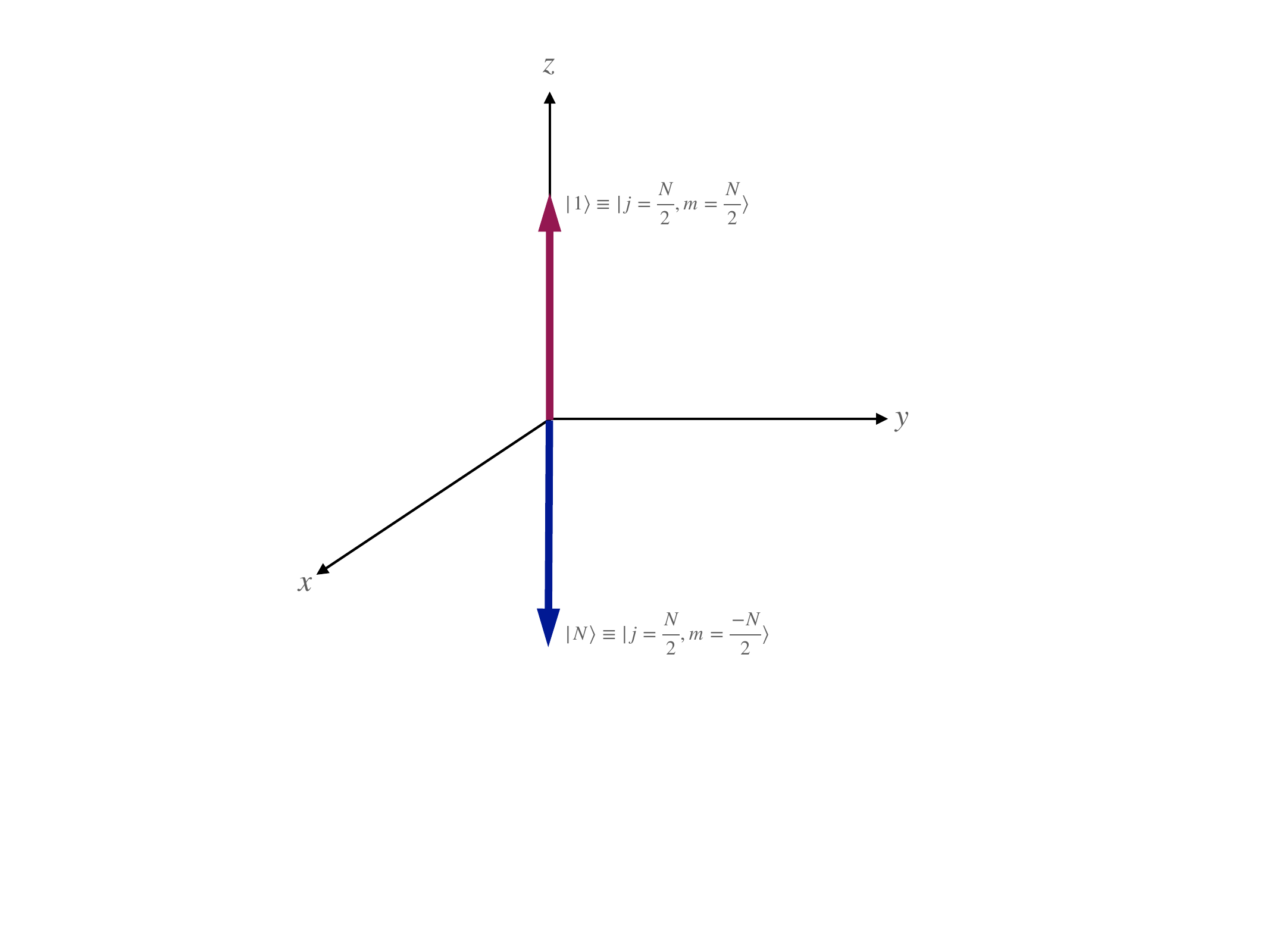}\vspace{-3cm}
	\caption{With the identification (\ref{nn}) the evolution operator $U$ perfectly transfers an excitation from the left hand side of the chain to the right hand side. }
	\label{zrot}
\end{figure}

	The analysis for short chains provides one with an insight or clue of how to proceed for PST on chains of arbitrary length. This scheme was suggested in the seminal work of \cite{ekert}. 	
	The main point is that the evolution operator should be such that at a specific time $T_0$, the state $|1,0,0,\cdots 0\ra$ to $|0,0,0,\cdots 1\ra$ modulo a phase. This phase can then be removed at the destination by an appropriate single qubit gate. Note that these two states are essentially states in a $2^n$ dimensional tensor product space $C_2^{\otimes n}$, i.e. 
	\be
	|1,0,0,\cdots 0\ra=|1\ra\otimes |0\ra\otimes |0\ra\otimes \cdots |0\ra, \h |0,0,0,\cdots 1\ra=|0\ra\otimes |0\ra\otimes |0\ra\otimes \cdots |1\ra.
	\ee
	However we can restrict ourselves to the one-particle sector of this space which is $N$ dimensional. In this space we can represent the above two states as 
	\be
|{ 1}\ra=	|1,0,0,\cdots 0\ra=\begin{pmatrix}1\\ 0\\ 0 \\ \cdot\\ \cdot \\ 0\end{pmatrix},\h 
|{ N}\ra=	|0,0,0,\cdots 1\ra=\begin{pmatrix}0\\ 0\\ 0 \\ \cdot\\ \cdot \\ 1\end{pmatrix}.
	\ee
	Note that the left hand side of these two equations are not transpose of the right hand sides. Although they  look similar, they point to quite different notations with different meanings. In view of the above representation, the evolution operator which is responsible for Perfect State Transfer (PST) should in fact rotate the state $|{\bf 1}\ra$ to the state $|{\bf N}\ra$ in one go, i.e. we should have 
	\be\label{rot1}
	U(T_0)\begin{pmatrix}1\\ 0\\ 0 \\ \cdot\\ \cdot \\ 0\end{pmatrix}=\begin{pmatrix}0\\ 0\\ 0 \\ \cdot\\ \cdot \\ 1\end{pmatrix},
	\ee
	in a proper time $T_0$ which can be determined. We note that the two states in (\ref{rot1}) are in fact the eigenstates of the angular momentum representation, then the operator $U(T_0)$ is nothing but a simple rotation around the $x$ axis. More concretely, in the angular momentum notation, where $$J_z|j,m\ra=m|j,m\ra,\h J^2|j,m\ra=j(j+1)|j,m\ra,$$ we can write 
		\be\label{nn}
|{ 1}\ra	= |j=\frac{N}{2},m=\frac{N}{2}\ra\h |{N}\ra= |j=\frac{N}{2},m=\frac{-N}{2}\ra.
	\ee
	Therefore the operator $U=e^{-i\pi J_x}$ which makes a rotation around the $x$ axis by an angle $\pi$, transforms the state 	$|j=\frac{N}{2},m=\frac{N}{2}\ra$ to the state 	$|j=\frac{N}{2},m=-\frac{N}{2}\ra$ in precise form, figure (\ref{zrot}). One can then make the statement that the Hamiltonian is $J_x$ and the time of evolution is $\pi$. The only thing which remains is to prove that the Hamiltonian $J_x$ can in fact represented by local interactions between spins in the Heisenberg chain and this is exactly what has been done in \cite{ekert}. We reproduce their argument in the following. Consider the Hamiltonian on a chain of length $N$. It is of the form
	\be
	H_N=-\frac{1}{2}\sum_k J_k (X_kX_{k+1}+Y_kY_{k+1})
	\ee
	where $X_k$ and $Y_k$ are Pauli matrices on site $k$ and $J_k=\sqrt{k(N-k)}$. \\
	Note that this kind of coupling has a mirror symmetry, that is 
	$$J_k=J_{N-k}.$$ Figure (\ref{nonuniform}) shows the pattern of couplings in a short chain. 
	Consider the basic local Hamiltonian between two neighboring spins 
	\be
	{\bf h}:=\frac{1}{2}(X\otimes X+Y\otimes Y).
	\ee
	It is easy to verify the following:
	\be
	{\bf h}|0,0\ra, \ \ \ {\bf h}|1,1\ra=0,\ \ \ {\bf h}|0,1\ra=|1,0\ra, \ \ \ {\bf h}|1,0\ra=|0,1\ra.
	\ee
This implies that the full Hamilotnian acts on any excitation as a hopping term as follows.
\be
H|n\ra=J_{n-1}|n-1\ra+J_n|n+1\ra.
\ee
If we now tune the coefficients $J_n$ as follows
\be
H|n\ra=\sqrt{(n-1)(N-n+1)}|n-1\ra+\sqrt{n(N-n)}|n+1\ra,
\ee
and remind ourselves of the correspondence (\ref{nn}), we see that $H$ is nothing but $H=\frac{1}{2}(J_++J_-)=J_x.$ This means that the evolution operator $U=e^{iHT}$ will rotate the state $|{\bf 1}\ra$ to $|{\bf N }\ra$, if we tune the time to be  $T=\pi$. 

In hindsight, we can now see the basic idea of the work of \cite{ekert}. The scheme works, because the eigenvalues of the Hamiltonian with mirror symmetric couplings, have the property that 
$e^{-i\lambda_nt_0}=(-1)^n$ for some specific time $t_0$. This allows that 
a single excitation placed at one end of the chain evolves to the opposite end with unit fidelity at a fixed time 
$t_0$.  The pattern of coupling being too restricted, a natural question is whether other more relaxed types of possibly more realistic couplings, i.e. those decaying with distance, also allow perfect state transfer or not. This question was answered in \cite{KayNN}, where the author showed that the basic requirements on the symmetry and eigenvalues, are still met in more general settings. By somehow reversing the question, the author introduced an iterative algorithm based on the inverse eigenvalue problem, which computes the parameters (e.g., magnetic fields, distances) needed to produce the given spectrum and hence achieve perfect state transfer in chains with more general couplings. 

\subsection{Perfect state transfer with uniform couplings}\label{pstkay}
\begin{figure}[H]
	\centering
	\includegraphics[width=13cm]{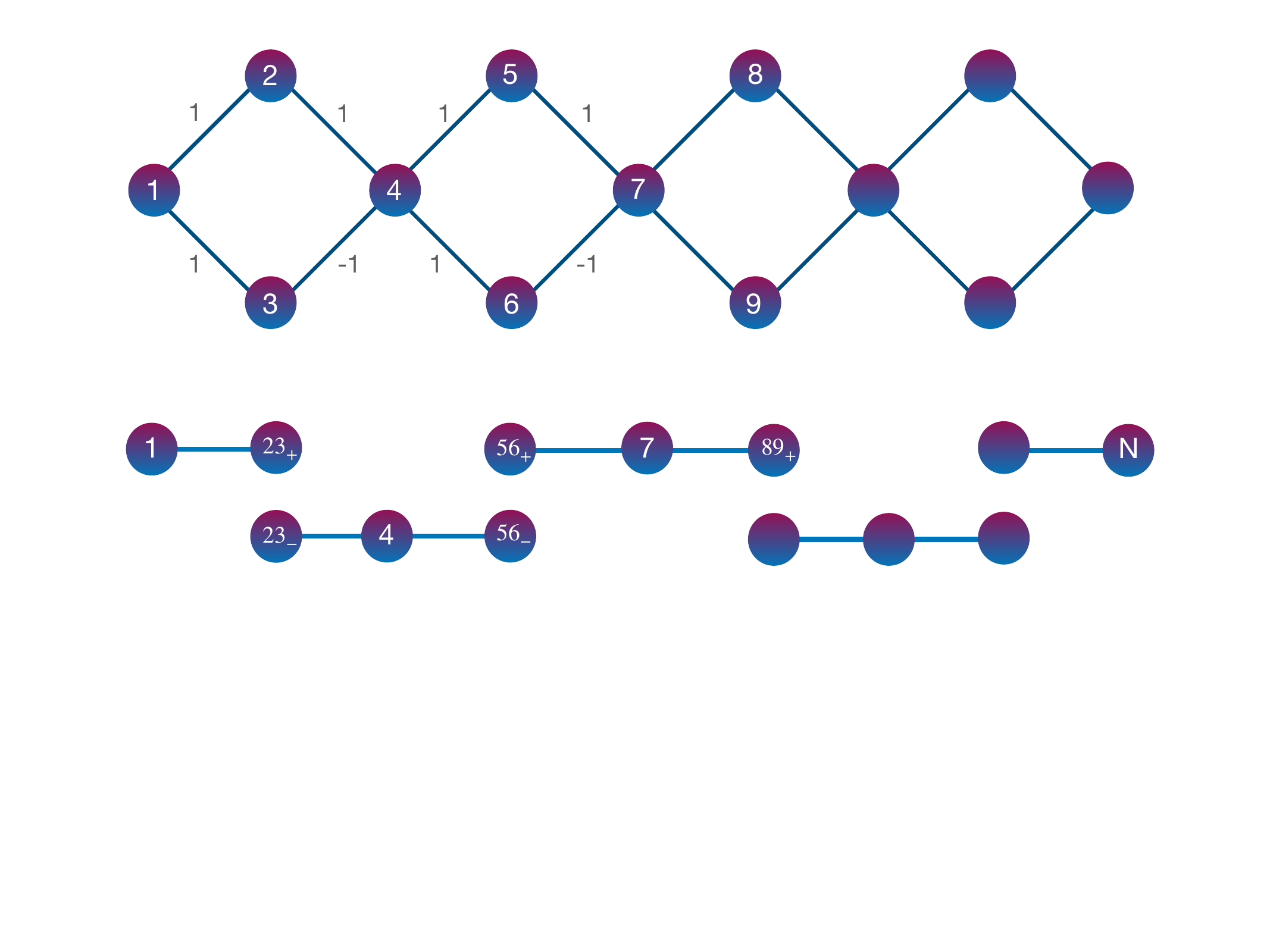}\vspace{-4cm}
	\caption{The quasi-one dimensional chain with uniform couplings for perfect transfer of quantum states. The figure below shows that the chain can be decomposed as a sequence of short chains of length 2 and 3.  }
	\label{1dchain}
\end{figure}

While perfect state transfer is achieved by the Heisenberg Hamiltonian, and it is possible in some platform to tune the couplings as required in this Hamiltonian, it is highly desirable if we could come up with a Hamiltonian with uniform couplings. It turned out that this is indeed possible, if we could pay a price. 
A technique for bypassing the need for engineering non-uniform interactions, was put forward in \cite{Kay} where the authors used a quasi-one dimensional chain in the form shown in figure (\ref{1dchain}). The price for this bypassing is that the natural dynamics of the chain should be accompanied by  global pulses at regular time intervals. The global nature of the pulse makes it experimentally feasible since there is no need for individual addressing of qubits. The crucial element in this proposal is the negative couplings in one of the legs in the diamond elements in this quasi-one dimensional chain. As we will see later on, this is related to the existence of the simplest Hadamard matrix, i.e. an orthogonal matrix  in two dimensions which reads 
$$H=\frac{1}{\sqrt{2}}\begin{pmatrix}1&1\\ 1&-1 \end{pmatrix}.$$
The scheme proposed in \cite{Kay} works as follows: 
The Hamiltonian is given by 
\begin{equation}\label{proto}
	H=\sum_{m,n} \frac{1}{2}K_{m,n}(X_mX_n+Y_mY_n).
\end{equation}
where $K_{m,n}$ are as shown in figure (\ref{1dchain}). Here we are denoting the Palui matrices by $X, Y$ and $Z$. First we note the following simple fact that on any two adjacent sites,
\be\label{perm}
\frac{1}{2}(X_mX_n+Y_mY_n)|\a,\beta\ra=|\beta,\a\ra
\ee
where $\a,\beta\in \{0,1\}$ and we are using the quantum computation notation $|0\ra=\begin{pmatrix}1\\ 0 \end{pmatrix}$ and  $|1\ra=\begin{pmatrix}0 \\ 1 \end{pmatrix}$. Therefore a state where all the spins are in the $|0\ra$ state is an eigenstate of this Hamiltonian and is invariant in time. 
One can also see that 
$$[H,\sum_n Z_n]=0,$$
which implies that the Hamiltonian conserves the total number of excitations. This means that we can denote the local Hamiltonian $H_{mn}$ by a simple permutation operator $P_{mn}$. Therefore, as explained above, the task of perfect transfer of an arbitrary state from left to right, reduces to the task of coherently transferring a single excitation $|1\ra$, i.e. from $|1\ra_1$ to $|1\ra_N$, where $N$ denotes the rightmost site in the chain. In view of (\ref{perm}), we can write the Hamiltonian as 
\be
H=P_{12}+P_{13}+P_{24}-P_{34}+P_{45}+P_{46}+P_{57}-P_{67}+\cdots.
\ee
Now, consider the one-particle sector spanned by the states $\{|1\ra,|2\ra, |3\ra, |4\ra, \cdots \}$. We can choose a new basis for this space as 
\be\label{new} \{|1\ra, |23_+\ra,|23_-\ra,  |45_+\ra, |45_-\ra, |67_+\ra, |67_-\ra, \ \ \ \cdots \},\ee where $$|23_{\pm}\ra=\frac{1}{\sqrt{2}}(|2\ra\pm |3\ra),\h |45_{\pm}\ra=\frac{1}{\sqrt{2}}(|4\ra\pm |5\ra), \ \ \ \ \cdots.$$ It is then easy to note that 
\be
(P_{12}+P_{13})|23_-\ra\propto(P_{12}+P_{13})(|2\ra- |3\ra)=|1\ra-|1\ra=0,
\ee
and
\be
(P_{24}-P_{34})|23_+\ra\propto (P_{24}-P_{34})(|2\ra+ |3\ra)=|4\ra-|4\ra=0.
\ee
This means that in the new basis (\ref{new}) the Hamiltonian has a block diagonal structure. This block diagonal structure is depicted in figure (\ref{1dchain}) which shows that the chain can be decomposed as a succession of uniform chains of length 2 and 3 which are decoupled from each other. 
We can find the expression of the Hamiltonian in each block as follows. We note that 
\be
(P_{12}+P_{13})|1\ra=|2\ra+|3\ra=\sqrt{2}|23_+\ra,
\ee and
\be
(P_{12}+P_{13})|23_+\ra=(P_{12}+P_{13})\frac{1}{\sqrt{2}}(|2\ra+|3\ra)=\sqrt{2}|1\ra.
\ee
Thus in the subspace spanned by $\{|1\ra, |23_+\ra\}$, which is decoupled from all other subspaces, the Hamiltonian acts as 
\be
H^{(2)}=\begin{pmatrix}0&\sqrt{2}\\ \sqrt{2}&0\end{pmatrix}.
\ee
Let us now determine the form of the Hamiltonian acting on a short chain of length $3$. Consider the subspace spanned by 
$\{|23_-\ra, |4\ra, |56_-\ra\}$. We have
\be
H^{(3)}=P_{24}-P_{34}+P_{45}+P_{46}.
\ee
By the same method as explained above, it is easily seen that 
\be
H^{(3)}|23_-\ra=\sqrt{2}|4\ra,\ \ \ H^{(3)}|4\ra=\sqrt{2}(|23_-\ra+|56_-\ra), \ \ \ H^{(3)}|46_-\ra=0. 
\ee
Thus the matrix form of the Hamiltonian in this subspace is given by
\be
H^{(3)}=\begin{pmatrix}0&\sqrt{2}&0\\ \sqrt{2}&0&\sqrt{2}\\ 0 & \sqrt{2}&0\end{pmatrix}.
\ee
Thus the tansfer of an excitation through the complete chain is reduced to its transfer through this succession of short chains of length 2 and 3 with uniform couplings, a transfer which we know is possible. The only problem is to transfer the particle from the right-hand-side of one chain to the left-hand-side of the next chain. This is achieved by applying a global pulse to the lower leg of the lattices at regular intervals of time. More concretely, consider now a state $|1\ra$. We now show how this state is transferred to the right hand side of the chain and become the state$|N\ra$ without leaving any trace of itself in the rest of the chain. 

A particle at site $1$ is transferred after a time lapse of $\tau=\frac{\pi}{2}$ to a linear superposition of particles at sites $2$ and $3$ in the form $\frac{1}{\sqrt{2}}(|2\ra+|3\ra)$ which we denote simply by $|23_+\ra$. A global $Z$ pulse to the lower chain in the form $Z=Z_2+Z_4+Z_6+\cdots$ now converts this state to $|23_-\ra:=\frac{1}{\sqrt{2}}(|2\ra-|3\ra)$. Note that the $Z$ pulse does not affect the empty sites and only negates the phase of the site which contains a particle. Now the particle at site $|23_-\ra$, after a time lapse of $\tau=\frac{\pi}{\sqrt{2}}$, is transferred perfectly to the end of the 3-chain in the form $|56_-\ra$ which under a global pulse  is converted to $|56_+\ra$. This state is again transferred via the 3-chain to $|89_+\ra$ and so on, until the particle arrives at the other end of the chain in perfect form. Therefore we need to apply regular global pulses to the lower chain at intervals of $\frac{\pi}{2}, \frac{\pi}{2}+\frac{\pi}{\sqrt{2}}, \frac{\pi}{2}+2\frac{\pi}{\sqrt{2}},\frac{\pi}{2}+3\frac{\pi}{\sqrt{2}},\cdots $ to achieve this perfect transfer in this quasi-one dimensional chain. A this point, the question arises whether or not perfect state transfer is also possible on two or even three dimensional arrays of qubits. After all the arrays of qubits which are assembled in platforms like optical lattices are two or three dimensional and it is highly desirable to find ways for perfect transfer in these lattices.\\

\section{Perfect routing of states in two and three-dimensional lattices}\label{pstme}
\begin{figure}
	\centerline{\includegraphics[scale=.26]{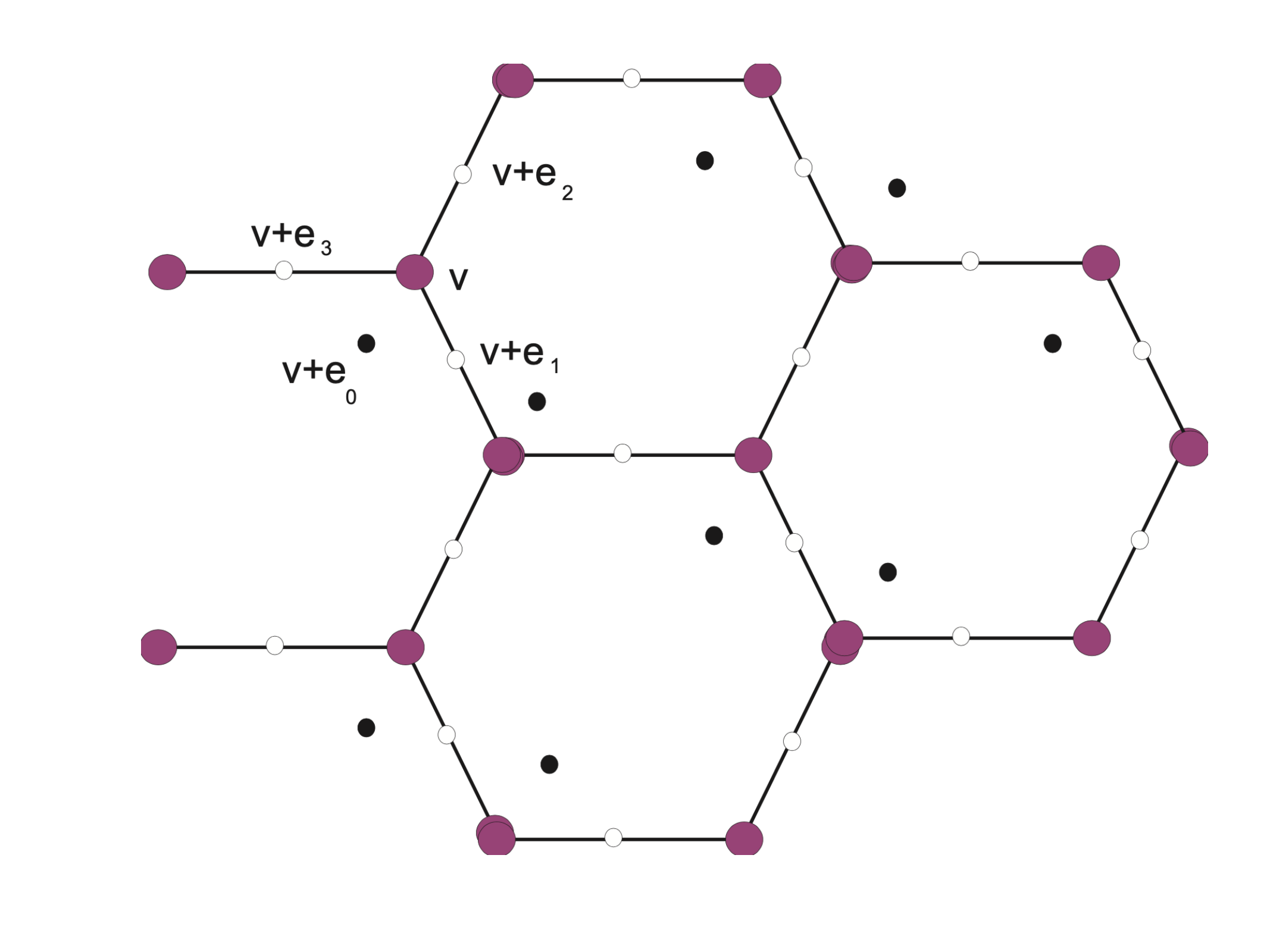}} \caption{(color
		online) The hexagonal plane. The RW heads are are the small black
		circles. The links accommodate qubit states (small white
		circles). The Hadamard switches (the big circles at vertices) are
		used to switch the qubit in different directions and the RW head
		(if necessary).} \label{Tiling}
\end{figure}

Many architectures of qubits in quantum computers are not one-dimensional. Atoms in optical lattices form two and three dimensional lattices, superconducting qubits are arranged in two dimensional lattices. Therefore it is important to invent schemes for perfect transfer of states in lattices beyond one dimension. On the other hand, while designing chains with non-uniform couplings may not be too difficult, it is certainly difficult to design such lattices with non-uniform couplings. Moreover, the argument of section (\ref{pstperfect}) on the analogy of rotation of spin $j$ states no longer works for two and three dimensional lattices. Therefore it is highly desirable to find a way for perfect transfer of states on these lattices without requiring non-uniform couplings. In \cite{sarmadi} a novel technique was invented for this purpose. The idea was inspired by the work of \cite{Kay}. In fact it was recognized in \cite{sarmadi} that it is the Hadamard matrix which plays a crucial role in the scheme of figure (\ref{1dchain}). 
It was therefore natural to ask whether or not the Hadamard matrix in the next dimension, namely dimension $4$ can play a similar role in higher than one dimensional lattices and surprizingly the answer was positive in a nice way. We remind the reader that a Hadamard matrix is a real orthogonal matrix in which all the absolute values of all the entries are equal. It is known that there is no Hadamard matrix in odd dimensions and There is only one Hadamard matrix in dimension 4. It is given by
\begin{equation}\label{Hadamard}
	{J}=\frac{1}{2}\left(\begin{array}{cccc}1&1&1&1\\ 1&1&-1&-1\\ 1&-1&1&-1\\
		1&-1&-1&1\end{array}\right).
\end{equation}
Inspired by this matrix and the work of \cite{Kay}, it was shown in \cite{sarmadi} that one can design a hexagonal Heisenberg lattices with 
uniform couplings to achieve perfect state transfer in two and three dimensional arrays of qubits. In the same way that the diamond like structure of cells in figure (\ref{1dchain}) is dicated by the two-dimensional Hadamard matrix, the hexagonal nature of the lattices in figure (\ref{Tiling}) is also dictated by the four dimensional Hadamard matrix. The reader may rightly ask in what sense the three neighbours of each site in figure (\ref{Tiling}) comply with the four-dimensional Hadamard matrix? This extra dimension is in fact a bona fide which allows us to use a fourth qubit as a read-write head near each site of the lattices. That is, one can upload any state from these read-write heads (shown as black colors filled circles in figure (\ref{Tiling})) to upload an arbitrary state $|\phi\ra$ to any point of the lattices and download it at any other site to the corresponding read-write head. In this respect, the extra dimension is a god-given aspect. In the following we explain this scheme in detail. \\

\textbf{The scheme} We start with the hexagonal lattices shown in
figure (\ref{Tiling}). Let $v$ denote a vertex of the lattices. On
the three links connected to this vertex, there are three qubits,
which we denote by $v+e_1$, $v+e_2$ and $v+e_3$. The vectors
$e_1$, $e_2$ and $e_3$ denote the three vectors directed along
the links connected to a vertex. A fourth qubit $v+e_0$, called
the read-write (RW) head is also connected to this vertex,
although the vector $v+e_0$  does not necessarily mean a vector
in the plane and is used only for uniformity of notation. The
Hamiltonian which governs the interaction on this system is of
the form
\begin{equation}\label{Hsum}
	H = \sum_{v}H_v
\end{equation}
where $H_v$ is the local Hamiltonian connecting each vertex to
its neighboring links and through these links to the other
vertices.\\

\begin{figure}
	\centerline{\includegraphics[scale=.26]{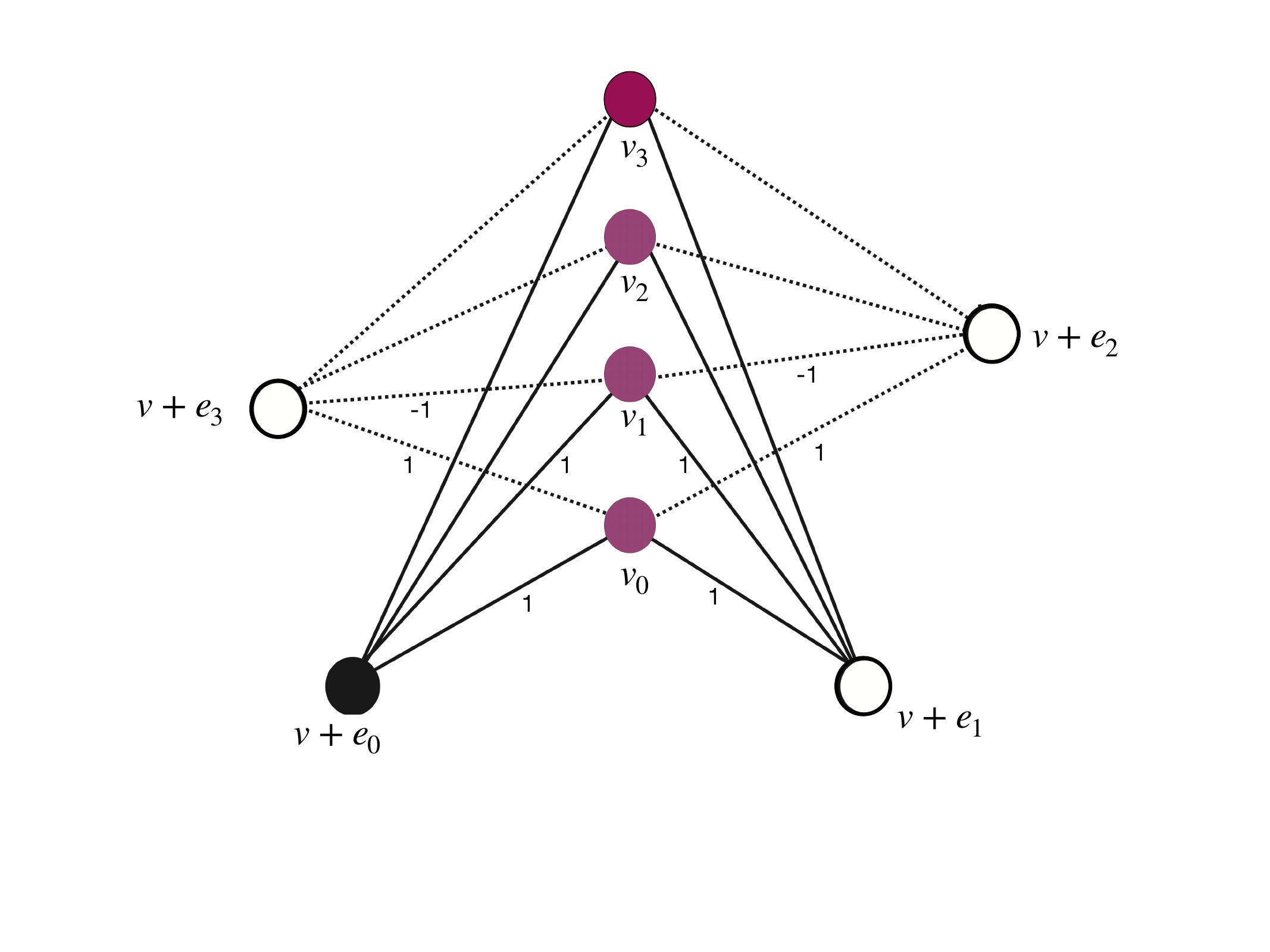}}\vspace{-1.6cm} \caption{(color
		online) The Hadamard switch. The four control qubits in the middle (the
		colored circles),  have an XY interaction with the four data qubits
		around according to the pattern of $\pm$ signs in the Hadamard
		matrix $J$ in (\ref{Hadamard}). For clarity of the figure, only the signs of the couplings of $v_0$ and $v_1$ to their numbers are shown. That there is no interaction
		between the central spins. The white circles are the qubits on the
		link and the black one is the (Read-Write) RW head. Only for clarity of the figure, the four control qubits are put on different control planes. In an actual architecture, these qubits need only be manipulated globally and simultaneously with no need for individual addressing. Hence they can be arranged in a way appropriate for that architecture.  }  \label{Joint}
\end{figure}

Now we describe the quantum system at each vertex. At each vertex
$v$ there are four qubits which we denote by $v_{_\a}$, i.e.
$v_{_0}, v_{_1}, v_{_2}, v_{_3}$. These qubits are arranged on
four different planes so that all the qubits with the same index
lie in one plane. In particular the qubits $v_0$ for different
$v$'s lie in the main hexagonal layer which accommodates the data qubits and the other qubits lie in
different planes which we call control planes to distinguish them
from the main hexagonal layer. At each vertex $v$, the  qubits $v_1, v_2$ and $v_3$ are called control qubits. 
 As we will see later, the only
control that we need is the possibility of applying uniform
magnetic field on each control plane. No operation be it unitary or measurement on any of the indivicual control bits is necessary. 
The interaction between all the
spins is a simple $XY$ interaction. Such an interaction conserves
the total z-component of spin and hence when the whole lattices is
initialized to the state $|0\ra$ (spin up), transfer of a
particle occurs in the single-particle sector. The local
Hamiltonian $H_v$ has the simple form
\begin{equation}\label{Hv}
	H_v = \sum_{\a,\b=0}^3 J^{\a\b} \left(X_{v_\a}X_{v+e_\b}+Y_{v_\a}Y_{v+e_\b}\right)
\end{equation}
where $J^{\a\b}$ are the entries of the Hadamard matrix in four
dimensions, namely
\begin{equation}\label{Hadamard}
	{J}=\frac{1}{2}\left(\begin{array}{cccc}1&1&1&1\\ 1&1&-1&-1\\ 1&-1&1&-1\\
		1&-1&-1&1\end{array}\right).
\end{equation}
More concretely, if we denote $\frac{1}{2}(X_iX_j+Y_iY_j)$ simply by ${\bf S}_i\cdot {\bf S}_j$, and refer to figure (\ref{Joint}), this Hamiltonian shows that there are spin-spin interactions of the form $\pm{\bf S}_i\cdot {\bf S}_j$ between all the nodes in this figure and signs are determined by the above matrix. For example the node $v_0$ is coupled with all its neighbors by $+$ sign, while the node $v_1$ is coupled with $+$ sign with the nodes $v+e_0 $ and $v+e_1$ and with $-$ sign with the nodes $v+e_2$ and $v+e_3$. 
We remind the reader that a Hadamard matrix is a symmetric real
orthogonal matrix all of whose entries have the same absolute
value. Such matrices exist only in certain special dimensions. We
will elaborate on the importance of this matrix for our scheme
later on. In the above matrix the rows and columns are numbered
from 0 to 3 from left to right and from top to bottom
respectively. Note that the vertex $v_0$ is connected with each of
the three links and also the RW head with equal couplings. We
call this structure, described by the Hamiltonian $H_v$ a
Hadamard switch. As we will see later, it can be used in a very
effective way for routing states through two and three
dimensional structures. Figure (\ref{Joint}) shows this switch.
In figure (\ref{Tiling}) these switches have been depicted as big
colored circles at vertices of the hexagonal lattices.\\

It can be readily verified that the local XY hamiltonian, on
spins $m$ and $n$, $H_{m,n}:=\frac{1}{2}(X_mX_n+Y_mY_n)$ has the
following simple action on the single excitation states, $
H_{m,n}|m\ra=|n\ra, \ \ H_{m,n}|n\ra=|m\ra, $ where
$H_{m,n}|p\ra=0$ for $p\ne m, \ n$. This means that $H_{m,n}$
when restricted to single particle subspace has the following
expression,
\begin{equation}\label{Hmnket}
	H_{m,n}=|m\ra\la n|+|n\ra\la m|.
\end{equation}
This allows us to rewrite the total Hamiltonian in the form
\begin{equation}\label{Hvket}
	H=\sum_{v,\a,\b}J^{\a,\b}(|v_\a\ra\la v+e_\b| + |v+e_\b\ra\la
	v_\a|).
\end{equation}
We now consider the four orthogonal states \ $
|\xi_v^{\beta}\ra:=\sum_{\b=0}^3 J^{\a,\b}|v_\a\ra,
$ 
that is
\begin{eqnarray}\label{allxi}
	|\xi_v^{0}\ra&:=&\frac{1}{2}(|v_0\ra+|v_1\ra+|v_2\ra+|v_3\ra)\cr
	|\xi_v^{1}\ra&:=&\frac{1}{2}(|v_0\ra+|v_1\ra-|v_2\ra-|v_3\ra)\cr
	|\xi_v^{2}\ra&:=&\frac{1}{2}(|v_0\ra-|v_1\ra+|v_2\ra-|v_3\ra)\cr
	|\xi_v^{3}\ra&:=&\frac{1}{2}(|v_0\ra-|v_1\ra-|v_2\ra+|v_3\ra).
\end{eqnarray}
It is here that the existence of Hadamard matrix in $4$ dimensions plays a crucial role, which allows to define these new orthogonal states all with uniform (up to signs) superpositions of the original states. 
The Hamiltonian can now be rewritten as
\begin{equation}\label{Hxi}
	H=\sum_{v,\beta}(|\xi_v^\a\ra\la v+e_\beta| + |v+e_\beta\ra\la \xi_v^\beta|).
\end{equation}
Thus in view of the hexagonal structure of the lattices shown in figure (\ref{Tiling}) and the structure of the Hadamard switch shown in figure (\ref{Joint}), we see that in this new basis for the control qubits, 
the Hamiltonian has been decomposed into direct sum of XY spin
chains with uniform couplings of length two and three. \\

We now
note that such chains are capable of perfect transfer of qubits
in times $t_0= \frac{\pi}{2}$ and $t_1=\frac{\pi}{\sqrt{2}}$
respectively.
\begin{figure}[t]
	\centering
	\includegraphics[width=12cm,height=10cm,angle=0]{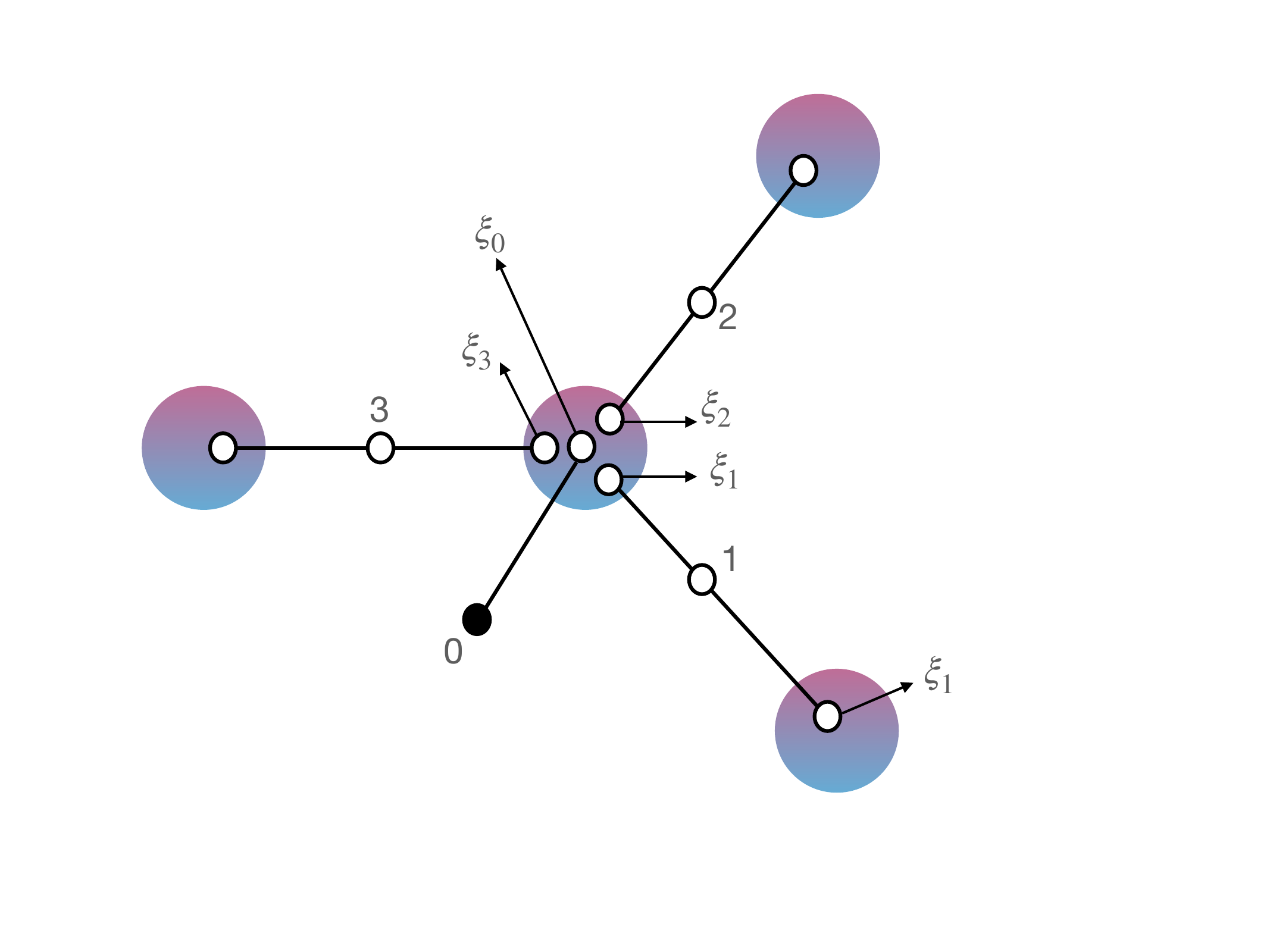}\vspace{-2cm}
	\caption{(color online) At each vertex, the control qubits $v_1, v_2, v_3$ and $v_0$ give their place to virtual qubits 
		$\xi_1, \xi_2, \xi_3$ and $\xi_0$. 		
		The local Hamiltonian $H_v$, when written in the
		basis $|\xi^\a\ra$, decomposes into perfect transfer effective XY
		chains of length two and three. The 2-chain transfers the particle
		between the RW head (the black circle) and the state $|\xi^0\ra$.
		Global control on the plane of spins 1, 2 or 3, changes the
		states $|\xi^\a\ra$ as desired. A state like $|\xi^1\ra$
		perfectly goes in the direction 1, while a state like $|\xi^2\ra$
		is stopped in this direction and goes only through direction 2.
		All the couplings on all the links are 1. The global pulses, $Z_1Z_2, Z_1Z_3$ and $Z_2Z_3$ displaces an excitation between these virtual sites which then is perfectly transferred to the end of its corresponding short chain.} \label{Headache}
\end{figure}
It is important to note that the states $|\xi_v^{\a}\ra$ are
turned into each other by global unitary operators. To see this clearly, we note that 
when the lattices is in the state $|v_1\ra$, it means that the only excitation which exists in the lattices is sitting on this single site of the plane no. 1.  
Let $Z^1$,
$Z^2$ and $Z^3$ be the Pauli operators $Z$ acting on all the spins in  planes 1, 2 and 3
respectively as follows, i.e. 
\be
Z^i=\prod_{ {\rm all\  Z's\  in\  plane\ i}}Z.
\ee
Such an operator which acts globally on the $i$-th plane, has the following action
\be
Z^i|v_i\ra\equiv Z_i|0,0,0,\cdots ,1_{v_i},\cdots 0,0,0\ra=-|0,0,0,\cdots ,1_{v_i},\cdots 0,0,0\ra=-|v_i\ra,  
\ee
regardless of $v$. Therefore when a $Z$ pulse acts on the $i-$th plane, it produces a minus sign only if there is an exitation on that plane, regardless of its place in that plane.  It is now readily seen that
\begin{eqnarray}\label{control}
	Z^1Z^2:|\xi^1_v\ra\rl |\xi^2_v\ra,\h |\xi^0_v\ra\rl |\xi^3_v\ra,\cr
	Z^1Z^3:|\xi^1_v\ra\rl |\xi^3_v\ra, \h |\xi^0_v\ra\rl |\xi^2_v\ra,\cr
	Z^2Z^3:|\xi^2_v\ra\rl |\xi^3_v\ra,\h |\xi^0_v\ra\rl |\xi^1_v\ra.
\end{eqnarray}
The crucial point is that the $Z^i$ operations can be applied
globally on all the qubits in the $i-th$ control plane, since on
the empty sites it has no effect and on an occupied site it has
the phase effect that we want. Therefore there is no need for
addressing single spins in each control plane, only the
possibility of access to each plane is required. Such a control
should be applied in a time much shorter than the time scale of
evolution of the Hamiltonian, namely $t_0$ and $t_1$. If we want
to route many particles at the same time, we have to control
different regions of control planes in accordance with the paths
of these particles. The control will be the same in the time
intervals when the paths become parallel.\\

Now a clear and very simple method for perfect state transfer in
the lattices emerges. A single particle  $\a|0\ra+\beta|1\ra$ is
uploaded to a given input head $v_{in}$. The part $\a|0\ra$ does
not evolve and indeed is ready for downloading at any output
head. We only have to transfer the single particle $|1\ra$ or excitation
which in view of our notation, has made the whole lattices to be
in the state $|v_{in}+e_0\ra$. After a time $t_0$, this state
evolves to $|\xi_{v_{in}}^0\ra$, i.e. the particle has moved to
the nearest  vertex $v_{in}$ in the form of a real wave $\xi^0$.
Once in a state $|\xi_{v}^0\ra$, we can make a global control
according to (\ref{control}) to switch this state to
either of the states $|\xi_v^i\ra$ ($i=1,2,3$) depending on the
direction we want to route the state. For example if we switch it
to $|\xi_{v}^1\ra$, then according to figure (\ref{Headache}),
after a time $t_1$, the state will be transferred perfectly to
the  vertex $v+e_1$ in the form $|\xi_{v+e_1}^1\ra$. Continuing in
this way we can move the state via any path that we like to any
other vertex say $v_{out}$, where the final state will be one of
the three states $|\xi_{v_{out}}^i\ra$, ($i=1,2,3$). Switching
this state to $|\xi_{v_{out}}^0\ra$ will move this state to the
nearest output head in the form $|v_{out}+e_0\ra$ where it will
be read off. The total time for routing is $2t_0+Nt_1$, where $N$
is the number of links which connect the input and output heads
along the chosen path. The sequence of control operations is very
simple. As seen from (\ref{control}) which is itself derived from (\ref{allxi}), we have to apply the following global pulses\\

$\bullet:$ for turning from direction $i$ to $j$ apply $Z^iZ^j$ abbreviated as $Z^{ij}$ in figure (\ref{Routing}),\\

$\bullet:$ For uploading and downloading a qubit to or from a RW
head to direction $i$ apply ${Z^i}$.\\

Except for uploading and downloading operations where a time
lapse of $t_0$ is needed all the other control operations are
applied at regular intervals of
time $t_1$.\\

\begin{figure}[H]
	\centerline{\includegraphics[scale=.28]{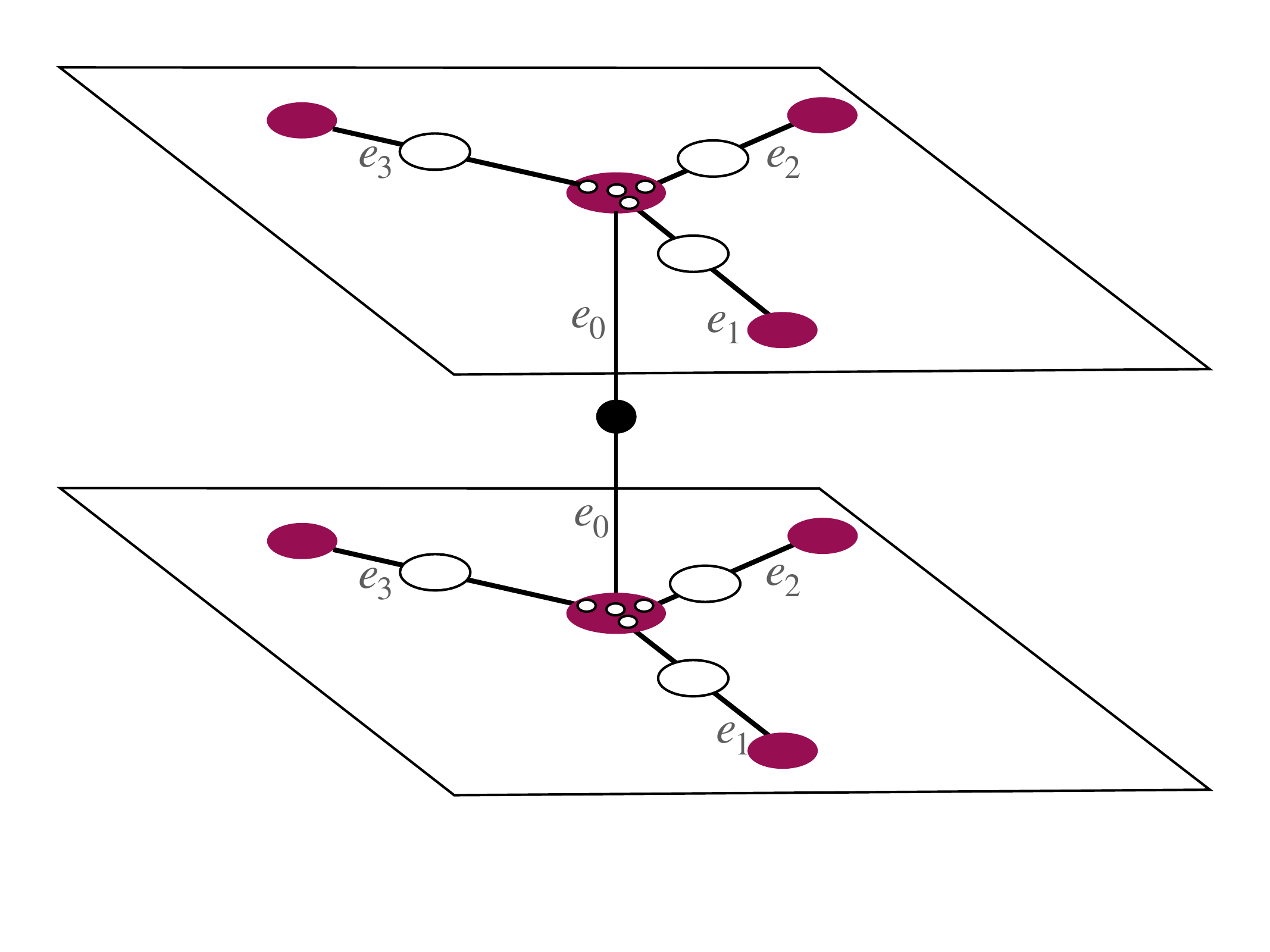}}\vspace{-1cm}
	\caption{(color online) The Hadamard switch can transfer a state
		between planes. The leg which was previously connected to RW
		heads is used to connect two switches in two different planes.
		The effective Hamiltonian on the 2-chain is now replaced by a
		3-chain connecting two planes.} \label{twolayers}
\end{figure}

\begin{figure}[H]
	\centerline{\includegraphics[scale=.38]{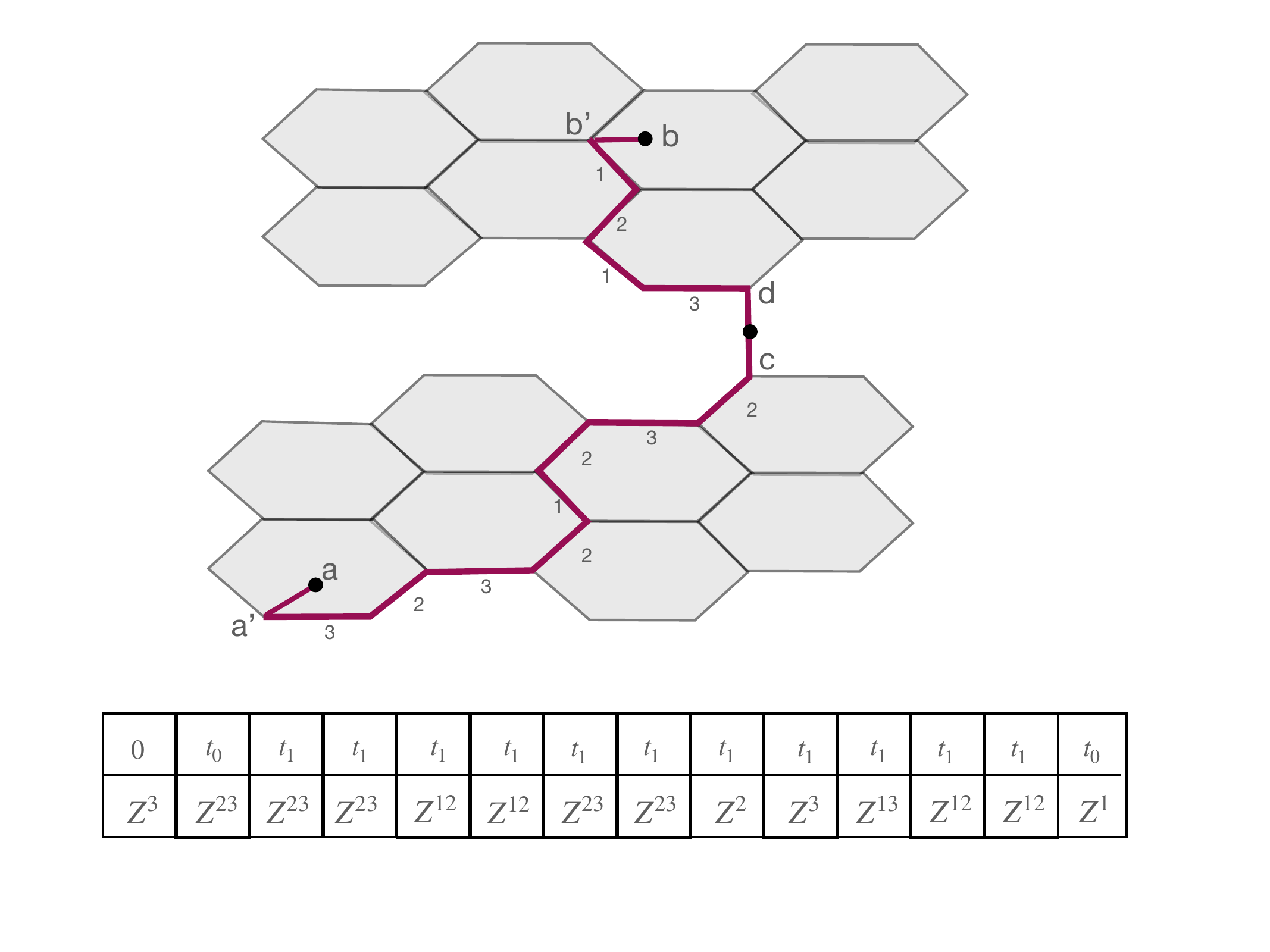}}\vspace{-1.2cm}
	\caption{(color online) The sequence of global pulses and the time difference between these pulses corresponding to the above route from node $a$ to node $b$ is as follows. $t_0=\frac{\pi}{{2}}$ and $t_1=\frac{\pi}{\sqrt{2}}$. $Z^{ij}$ stands for $Z^iZ^j$. To upload a particle (i.e. an excitation) from $a$ to $a'$ at the begining of a link $e_3$ takes a time $t_0$ and a pulse $Z^3$, since this is a short chain of length 2. Similarly downloading the particle from site $b'$ at the end of a link $e_2$ to $b$ takes a time $t_0$ and a pulse $Z^1$. However to guide the particle from the point $c$ at the end of a link $e_2$ in the lower plane, to the point $d$ at the begining of a link $e_3$ on the upper plane takes two consecutive pulses $Z$ and then $Z^3$ and a time $t_1$, since $cd$ is a long chain of three spins. See figure (\ref{twolayers}).
	} \label{Routing}
\end{figure}

\textbf{Perfect transfer in three dimensions} The Hadamard switch
can be used in another way for achieving perfect state transfer
in three dimensional structures. Figure (\ref{Joint}) shows a
Hadamard switch connecting two hexagonal planes. Such planes can
be joined by any number of switches. The number and positions of
Hadamard switches are determined to optimize the accessibility of
all the heads in the two planes by shortest possible paths. When
used in this new way, the RW head gives its role to the qubit on
the link which joins the two planes. For example when the two
planes are joined to each other at points $x_1$, and $x_2$ on the
two planes (figure \ref{twolayers}), the effective Hamiltonian
for the states $|x+e_0\ra$ on the joining link and the states
$|\xi_{x_1}^{0}\ra $ on plane 1 and $|\xi_{x_2}^{0}\ra $ on the
upper plane is nothing but a perfect XY 3-chain. This effective
Hamiltonian, transforms the state $|\xi^0\ra$ perfectly between
the two planes. This time we should wait for time $t_1$ instead
of $t_0$. Therefore we route a particle within each plane as
before and bring it to the position of the nearest switch where by
appropriate control we move it to another plane and continue
there.\\

\textbf{Robustness} Like the system in \cite{Kay}, our
scheme is also robust (in a limited sense described below) against imperfections in the lattices up to
a threshold. By this limited robustness, we mean that if certain number of Hadamard switches are not working properly, due to defects in their couplings, we can  easily route them around. Needless to say, this robustness does not cover imperfections in all couplings of the lattice. For the simple one dimensional scheme, this has been studied in \cite{benattime}. 
 By the threshold here, we mean that the number of switches in the lattice which are not working must be low enough to allow alternative routes. 
 Moreover due to the very
simple nature of the switches and their control, if there is any
delay in the control operations, we know exactly on which switch
and which link of the switch the particle is waiting. This is also true if a connection (or the Hadamard switch) between two planes is not working. In that case we can route away the particle from that particular switch and guide it via  a different link. This property  is due
to the fact that under the intrinsic dynamics of the 2 and 3
chains in figure (\ref{Headache}), any excitation just goes
back and forth between the endpoints of a chain.\\

	\section{Discussion} 
	We have revisited the problem of quantum state transfer in qubit arrays with a detailed and pedagogical approach. Our focus is on perfect state transfer in two- and three-dimensional lattices, where quantum states are transferred with perfect fidelity between stacked planes. The use of uniform ferromagnetic or antiferromagnetic couplings between spins requires additional control qubits, to which control pulses are applied simultaneously without individual addressing. These control qubits do not necessarily need to be placed on different planes; their optimal arrangement depends on the chip's architecture.	
	A key point is that, although we use the notation, concepts, and diagrams of spin chains, the actual qubits are superconducting, with interactions that can be effectively modeled by XY spin couplings under certain conditions. Given the rapid advancements in quantum chips based on semiconductor qubits and the push for miniaturization—potentially through multi-layer technology—our work may have practical applications.\\
	
	For brevity, some aspects of the problem have been left out. Specifically, we have not addressed perfect state transfer in spin-
	one chains \cite{asoudehme}, the impact of thermal noise  \cite{bayatme}, or imperfections in timing of pulses on input-output fidelity \cite{benattime}. We have left out the use of quasi-one dimensional chain of figure (\ref{1dchain}) for quantum state generation \cite{moradi} and the construction of static quantum circuits by using quantum spin chains.  \cite{seifnashri}.\\

	\textbf{Acknowledgement} This research was supported in part by Iran National Science Foundation under Grant No.4022322. M.A. would like to thank the support of the office of research in Azad University. \\
	
	\textbf{Data Availability Statement:} No Data associated in the manuscript.

\end{document}